\documentclass{article}
\usepackage{natbib}
\usepackage[utf8]{inputenc} 
\usepackage[T1]{fontenc}    
\usepackage{hyperref}       
\usepackage{url}            
\usepackage{booktabs}       
\usepackage{amsfonts}       
\usepackage{nicefrac}       
\usepackage{microtype}      
\usepackage{float}
\usepackage{subcaption}
\usepackage{amsmath}
\usepackage{graphicx}
\usepackage[export]{adjustbox}

\usepackage{mwe,lipsum}
\makeatletter\ifdefined\Hy@StartlinkName
\def\addOneNestingLevelStartLink{%
  \gdef\Hy@StartlinkName##1##2{%
    \sbox0{\Hy@StartlinkNameOrig{##1}{##2}}\usebox0
    \global\let\Hy@StartlinkName\Hy@StartlinkNameOrig%
  }%
}
\def\addOneNestingLevelEndLink{%
  \gdef\pdfendlink{%
    \sbox0{\pdfendlinkOrig}\usebox0%
    \global\let\pdfendlink\pdfendlinkOrig%
  }%
}
\let\Hy@StartlinkNameOrig\Hy@StartlinkName
\let\pdfendlinkOrig\pdfendlink
\else
\let\addOneNestingLevelStartLink\relax
\let\addOneNestingLevelEndLink\relax
\fi

\title{SMOGS: Social Network Metrics of Game Success}

\author{Fan Bu \and Sonia Xu \and Katherine Heller \and Alexander Volfovsky}
\date{Department of Statistical Science\\
\vspace{0.1in}
Duke University}

\usepackage{geometry}
\begin{document}

\maketitle
%
\begin{abstract}
This paper develops metrics from a social network perspective that are directly translatable to the outcome of a basketball game. We extend a state-of-the-art multi-resolution stochastic process approach to modeling basketball by modeling passes between teammates as directed dynamic relational links on a network and introduce multiplicative latent factors to study higher-order patterns in players' interactions that distinguish a successful game from a loss. Parameters are estimated using a Markov chain Monte Carlo sampler. Results in simulation experiments suggest that the sampling scheme is effective in recovering the parameters. We then apply the model to the first high-resolution optical tracking dataset collected in college basketball games. The learned latent factors demonstrate significant differences between players' passing and receiving tendencies in a loss than those in a win. The model is applicable to team sports other than basketball, as well as other time-varying network observations. 
\end{abstract}

\section{Introduction}
The game of basketball is a sport played between two teams of five players, in which the game is won by seeing which team can obtain the most baskets in the time alloted. Starting from the onset, a basketball game continuously evolves in time and space and constitutes a constant flow of player movements, interactions, and decision making that contribute to the game outcome. As a team sport, basketball requires the collaboration of the players to successfully bring the ball to the basket, and such collaboration relies heavily on passing the ball between teammates. Understanding and evaluating the decisions made by players on whether to pass, when to pass, and whom to pass to is an ongoing challenge in sports analytics. 

Since passes between teammates can be modeled as interactions within a network, a variety of previous methods have taken a network approach to studying passing sequences. \cite{fewell2012basketball} focus on exploratory network analysis to explain key facets of the game like key players and specific team play styles, where the roles of different players on a team level are explained through weighted graphs of passing frequencies. \cite{gudmundsson2017spatio} calculate rebound probability with spatial coordinates to measure team and player performance via graph theory. \cite{xin2017continuous} implement a continuous-time stochastic block model to cluster players based on passing networks. 

Although basketball games were traditionally analyzed in a discretized manner based on ``box score'' statistics for forecasting \citep{hollinger2005pro} and player evaluation \citep{omidiran2011new}, the installment of optical tracking systems in professional basketball arenas in 2013 allows more detailed statistical analyses. High-resolution spatial and temporal information has been leveraged to model the courses of games and characterize basketball strategies that were overlooked in low-resolution analyses. \cite{miller2014factorized} summarize player shot locations as low dimensional spatial bases by smoothing empirical shot locations through non-negative matrix factorization (NMF). The spatial bases for each player translate well in determining a player's position on the team. \cite{pelechrinis2017thoops} use tensor decomposition to create a weighted shot chart from a 12-dimensional representation of each player. \cite{franks2015characterizing} characterize the spatial structure of defensive basketball play and quantitatively evaluate the guarding choices and movements of defending players. \cite{cervone2016multiresolution} 
develop a multi-resolution stochastic process model to calculate the expected points the offense will score in a possession conditional on the evolution of the game up to a time point. 


    
In recent years, an NCAA Division I basketball team partnered with SportsVu to install optical tracking systems in their home stadium, collecting the first high-resolution spatio-temporal dataset of college-level basketball games (a detailed description of this dataset is given in section 4.2). This paper aspires to evaluate player interactions and develop metrics for game success that are applicable to (but not restricted to) collegiate basketball settings. We build on the work conducted by \cite{cervone2016multiresolution} by modeling passes from a network perspective and introducing multiplicative latent factors to study players' passing choices and preferences. These multiplicative effects provide a novel assessment of the efficacy and effectiveness of the passing game of a team.

Treating a pass from player $i$ to player $j$ at time $t$ as a dynamic relational link from $i$ to $j$, the observations of whether and between whom a pass is made in a basketball possession conditional on the spatio-temporal evolution of the possession up to time $t$ are analogous to repetitive observations of relational ties on networks. In the last two decades, some authors have used non-additive random effects on top of fixed covariate effects to model nodal links on networks. \cite{nowicki2001estimation} assume that the probability of a link between two nodes depends on the shared membership in a collection of latent classes. The more general class of latent space models maps nodal characteristics onto an unobserved social space with ties depending on the similarity between actors within the latent space \cite{hoff2002latent,hoff2005bilinear}. This class of models has been extended to allow heterogeneous additive and multiplicative sender- and receiver-effects \citep{hoff2009multiplicative,hoff2013likelihoods} as well as to dynamic networks \citep{durante2014nonparametric,sewell2015latent}. 
This work extends the class of latent factor network models by modeling link occurrences as non-homogeneous Poisson processes on a dynamic network in the complex spatio-temporal setting of basketball games.



The remainder of the paper is organized as follows. In the next section we provide an overview of the key aspects of both the stochastic model by \cite{cervone2016multiresolution} and a latent factor social network model, in a basketball setting. In section 3, we present our novel latent factor stochastic passing model in detail and discuss our parameter estimation procedure. Section 4 presents experimental results on synthetic datasets and real optical basketball tracking data, and lastly, the conclusion is in section 5. 

\section{Background}



\subsection{Multiresolution stochastic process model}

The model introduced by \cite{cervone2016multiresolution} begins with a coarsened view of a basketball possession. At any time point, the going-ons in a game fall into one of the three types of states: a possession state, a transition state, and an end state. The ball does not change hands in a possession state, and thus this state can be modeled by the micro movements of players. An end state, as suggested by the name, simply represents the end of the possession via points (0, 2, or 3) earned by the offense. It is within a transition state that the dynamics of a basketball game changes qualitatively: a transition can be a pass, a shot attempt, or a turnover, after which either the ball carrier changes or the possession ends. Based on the assumption that, given the ending state of a transition, a future possession is conditionally independent of the history up to the beginning of that transition, modeling the occurrence and end state of a transition is essential to predict the outcome of a possession well. 

Among the three transitions (pass, shot attempt, turnover), a shot attempt results in either a made shot or a failed shot, a turnover leads to a change of possession, but a pass has four potential outcomes corresponding to the four other teammates as potential receivers, which depend on various spatio-temporal factors. More specifically, assuming player $i$ possesses the ball at time $t$, let $\theta_{i,j}(t)$ be the hazard for the occurrence of a pass to teammate $j$ in $(t,t+\epsilon]$,
\begin{equation}
\theta_{i,j}(t) = \lim_{\epsilon \rightarrow 0+} \frac{\mathbb{P}(\{\text{$i$ passes to $j$ in $(t,t+\epsilon]$}\}|\mathcal{H}(t))}{\epsilon},
\label{eq:epv-hazard}
\end{equation}
where $\mathcal{H}(t)$ denotes the history of the game up to time $t$. Further assume the hazard is log-linear,

\begin{equation}
\log(\theta_{i,j}(t)) = W_{i,j}(t)^T\eta_{i,j} + \xi_{i,j}(\mathbf{s}_i(t))+ \tilde{\xi}_{i,j}(\mathbf{s}_j(t)),
\label{eq:epv-loglinear}
\end{equation}
where $W_{i,j}(t)$ is a time-varying covariate vector, $\eta_{i,j}$ is the corresponding coefficient vector, and $\xi_{i,j}(\mathbf{s}_i(t))$ and $\tilde{\xi}_{i,j}(\mathbf{s}_j(t))$ are functions that map the ball carrier's location and the potential receiver's location on the court to additive spatial effects (a detailed description follows in section 3).

\subsection{Multiplicative latent factor model}
Observations on a social network can be characterized by an $n \times n$ matrix $\mathbf{Y}=\{y_{ij}\}$ where $y_{ij}$ is a binary variable representing the existence of a link from node $i$ to node $j$. Given a set of covariate vectors $\mathbf{X}=\{\mathbf{x_{ij}}\}$, a common approach to modeling the association between $\mathbf{Y}$ and $\mathbf{X}$ that also accounts for unobserved dependencies is to use random effects models,
\begin{equation}
\log\text{odds}(y_{ij}=1) = \theta_{ij} = \beta'\mathbf{x}_{ij} + z_{ij},
\label{eq:network-logit}
\end{equation}
where $z_{ij}$ is the unobserved random effect of pair $(i,j)$, and the $z_{ij}$'s are not necessarily independent so as to capture potential dependencies in the relational observations. \cite{hoff2005bilinear} and \cite{hoff2009multiplicative} motivate multiplicative effects on top of additive effects to model $z_{ij}$ to represent higher-order network structure,
\begin{equation}
z_{ij} = a_i + b_j + u_i'v_j + \epsilon_{ij},
\label{eq:add-mult-effects}
\end{equation}
where $a_i$ and $b_j$ are additive row-specific and column-specific effects that respectively represent the proclivity of player $i$ to pass the ball and the popularity of player $j$ as a receiver of the ball, $u_i$ and $v_j$ are multiplicative latent space factors, and the $\epsilon_{ij}$'s are independent random errors.

A variation of such model, an additive and multiplicative effects (AME) model was previously fit on aggregated passing networks in possessions using a subset of the college basketball optical tracking data to model the passing structure of the Division I basketball team. Let $y_{ij}$ be the indicator of whether player $i$ passes to player $j$ in a possession, and assume that


\begin{equation}
\log\text{odds}(y_{ij}=1) =  \beta_{d}x_{d} + r_{i} + s_{j} + u_{i}^{T}v_{j} + \epsilon_{ij},
\end{equation}
where $s_j=\beta_jx_j + b_j$ represent additive sender effects and $r_i = \beta_i x_i + a_i$ represent additive receiver effects. 
The dyadic features are represented by $\beta_{d}x_{d}$, 
and the multiplicative sender and receiver effects by $u_{i}^{T}v_{j}$. The additive sender and receiver effects include row- and column- specific effects ($a_{i}$, $b_{j}$) and row- and column- specific nodal features ($\beta_{i}$, $\beta_{j}$).

For this model, the dyadic features include indicators for shared basketball position, shared height, shared weight, and shared college class between players. Nodal features include binary variables of whether a player was in a previous possession, whether a player is in a current possession,and points earned per game by a player.   

\begin{figure}[H]
	\centering
	\includegraphics[width=.45\textwidth]{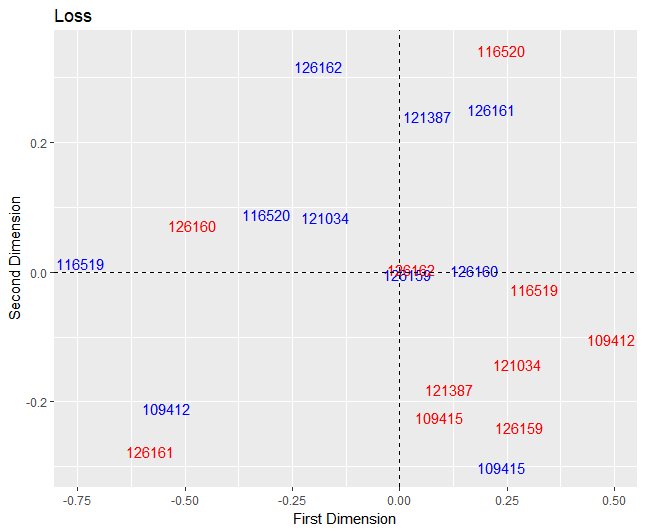}
	\hspace{0.1in}
	\includegraphics[width=.45\textwidth]{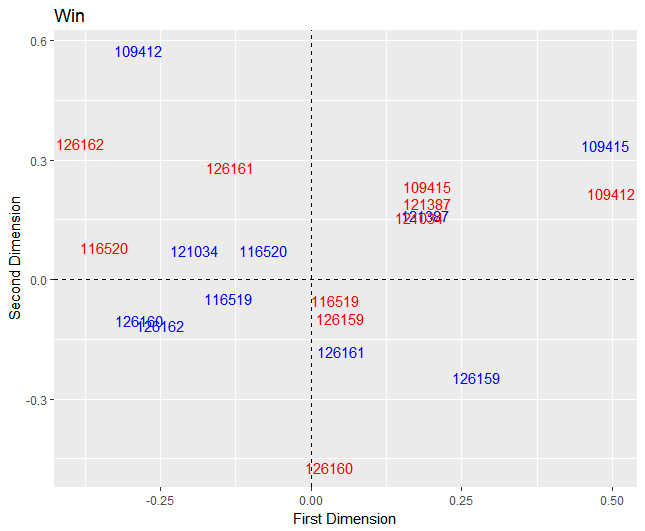}
	\caption{Learned multiplicative sender-specific effects (in blue) and receiver-specific effects (in red) by the additive and multiplicative effects model in a lost (left) game and a won game (right).}
	\label{fig:short-uv}
\end{figure}

Parameters are inferred for each game separately using the Markov chain Monte Carlo sampling algorithm suggested by \cite{hoff2008modeling} implemented in \cite{hoff2014amen}. Figure \ref{fig:short-uv} presents the posterior means of the multiplicative sender (passer) and receiver effects for a win and a loss. 

Looking at the posterior means for the multiplicative sender-specific (blue) and receiver-specific (red) effects, players who are in the same quadrant and differing colors are more likely to interact. Players that lie on the origin either pass or receive passes at the same rate from all other players. Operationally, the only players that could exhibit such behavior are point guards because of their neutrality in passing and receiving the ball from other players. In the loss plot, the two players who lie on the origin were not the point guards for this Division I Basketball team: Player 126162 (in the red) is a post player, and Player 126159 (in the blue) is a guard, suggesting non-standard playing. We can also evaluate the overall passing in the game using these plots. In the win we see the main starters for the team  clustered together in the left quadrant suggesting that they move the ball among themselves with high probability. Such behavior is not observed in the loss plot and in fact we see that ball movement was likely to be fragmented: there is only one player who is likely to receive passes from the passers in the upper left quadrant and only one player who is likely to pass to the others in the lower left quadrant. On the other hand, we see a substantially more even distribution of high pass probabilities in the win plot. 

\section{Model}
The findings of the AME model presented in section 2.2 demonstrate that the learned latent factors are directly translatable to players' passing patterns which are distinctive between a win and a loss, and that latent factor models help uncover the network characteristics in passing which are predictive of basketball game outcomes. 
In this section we introduce latent factors in the stochastic process setting of basketball passes and state our full model.
\subsection{Model formulation}
Let $Y_{i,j}(t)$ denote the event that the ball carrier $i$ passes the ball to teammate $j$ during the time period $(t,t+\epsilon]$ in game $g$, and let $\theta_{i,j}(t)$ be the risk of $Y_{i,j}(t)$ given $\mathcal{H}(t)$, the history up to time $t$ in game $g$, 
\begin{equation}
\theta_{i,j}(t) = \lim_{\epsilon \rightarrow 0+} \frac{\mathbb{P}(Y_{i,j}(t)|\mathcal{H}(t))}{\epsilon}.
\end{equation}
Assume that 
\begin{equation}
\log(\theta_{i,j}(t)) = W_{i,j}(t)^T\eta_{i,j} + \xi_{i,j}(\mathbf{s}_i(t))+ \tilde{\xi}_{i,j}(\mathbf{s}_j(t))+ z_{i,j,g}(t),
\label{eq:log-risk}
\end{equation}
and the latent factor 
\begin{equation}
z_{i,j,g}(t) = u_{i,g}^Tv_{j,g} + \epsilon_{i,j}(t).
\label{eq:latent-svd}
\end{equation}

In Eq.\ref{eq:log-risk} $W_{i,j}(t)$ is a 5-dimensional time-varying covariate vector, including a constant 1 representing the baseline passing rate, an indicator of whether player $i$ has started dribbling at time $t$, the log-transformed distance between player $i$ and his nearest defender, $j$'s rank of closeness to $i$ (from 1 to 4, with 1 indicating the closest teammate), a numeric evaluation of how open the passing route is from $i$ to $j$ (a metric introduced by \citep{cervone2016multiresolution}), while $\xi_{i,j}$ maps player $i$'s location on the half court to an additive spatial effect of passing off the ball, and $\tilde{\xi}_{i,j}$ maps a player $j$'s location on the half court to an additive spatial effect of receiving a pass from $i$ based on $j$'s basketball position, with $\epsilon_{i,j}(t)$'s as independent standard normal errors. In Eq.\ref{eq:latent-svd},  $u_{i,g}$ and $v_{j,g}$ are $R$-dimensional vectors 
representing sender-specific and receiver-specific attributes of ball carrier $i$ and teammate $j$ mapped onto an $R$-dimensional latent space.

Furthermore, let 
\begin{equation}
\begin{aligned}
\xi_{i,j}(\mathbf{s}) &= \gamma_{i,j} \bar{\xi}_{i}(\mathbf{s}),\\
\tilde{\xi}_{i,j}(\mathbf{s}) &= \tilde{\gamma}_{i,j}\bar{\tilde{\xi}}_{i,\text{pos}(j)}(\mathbf{s}),
\end{aligned}
\label{eq:spatial-func}
\end{equation}
where $\text{pos}(j)$ denotes player $j$'s basketball position, and $\int_{\mathcal{S}}\bar{\xi}_{i,j}(\mathbf{s})d\mathbf{s} =\int_{\mathcal{S}}\bar{\tilde{\xi}}_{i,j}(\mathbf{s})d\mathbf{s} = 1$, with $\mathcal{S}$ as the half court and $\mathbf{s}$ as a pair of coordinates corresponding to a location on $\mathcal{S}$. Therefore, setting $X_{i,j}(t) = (W_{i,j}(t)^T,\bar{\xi}_{i,j}(\mathbf{s}_i(t)),\bar{\tilde{\xi}}_{i,j}(\mathbf{s}_j(t)))^T$ and $\beta_{i,j} = (\eta_{i,j}^T,\gamma_{i,j},\tilde{\gamma}_{i,j})^T$, Eq.\ref{eq:log-risk} becomes
\begin{equation}
\log(\theta_{i,j}(t)) = X_{i,j}(t)^T\beta_{i,j} + u_{i,g}^Tv_{j,g} + \epsilon_{i,j}(t).
\end{equation}

We would like to emphasize that the model formulation is not only an extension of Eq.\ref{eq:epv-loglinear}, but also an extension of Eq.\ref{eq:network-logit} and \ref{eq:add-mult-effects}. A hierarchical structure in the multiplicative latent factors is introduced to allow differing sender-specific and receiver-specific effects in different games, and here we model the time-varying risk (intensity function) of a non-homogeneous spatio-temporal Poisson process rather than the probability of binary links using logistic regression. 

\subsection{Parameter estimation}
Conditioning on all the covariate vectors $X_{i,j}(t)$ regarding $n$ players in $G$ games in total, the unknown quantities of the model are
\begin{itemize}
\item $\Theta=\{\theta_{i,j}(t)\}$, the set of risks;
\item $\mathbf{\beta}=\{\beta_{i,j}\}$, the set of coefficients for all player pairs;
\item $\mathbf{U}=\{U_g\}$, the set of $n \times R$ matrices, with each row representing the sender-specific effect of a player in a game, and $\mathbf{V}=\{V_g\}$, the set of $n \times R$ matrices, with each row representing the receiver-specific effect of a player in a game.
\end{itemize}
Parameters can be estimated in a Bayesian fashion, where a joint posterior distribution of the parameters is obtained via the Bayesian rule given a reference prior, and quantities of interest such as parameter estimates and confidence intervals are approximated through Markov chain Monte Carlo sampling. More specifically, a Metropolis-within-Gibbs sampling scheme is adopted here. Given all the observations $\mathbf{Y}=\{Y_{i,j}(t)\}$ and a set of initial values $\phi^{(0)}=\{\Theta^{(0)},\mathbf{\beta}^{(0)},\mathbf{U}^{(0)},\mathbf{V}^{(0)}\}$, a sequence of parameter samples $\{\phi^{(s)}\}$ can be iteratively generated from the full conditional distributions of the parameters. In each iteration, based on the latest parameter sample $\phi^{(s-1)}$, a new sample $\phi^{(s)}$ is acquired through the following steps:
\begin{enumerate}
\item For each pair $(i,j)$, $1\leq i \neq j \leq n$, sample $\beta_{i,j}^{(s)}$ from $p(\beta_{i,j}|\Theta^{(s-1)},\mathbf{U}^{(s-1)},\mathbf{V}^{(s-1)})$;
\item For game $g=1,\dots,G$ and $r=1,\ldots,R$,
\begin{enumerate}
\item sample $U_g[,r]^{(s)}$ from $p(U_g[,r]|\Theta^{(s-1)},\mathbf{\beta}^{(s-1)},U_g[,-r]^{(s-1)},V_g^{(s-1)})$;
\item sample $V_g[,r]^{(s)}$ from $p(V_g[,r]|\Theta^{(s-1)},\mathbf{\beta}^{(s-1)},U_g^{(s-1)},V_g[,-r]^{(s-1)})$;
\end{enumerate}
\item For each pair $(i,j)$ and time point $t$, set $\theta_{i,j}^*(t) = X_{i,j}(t)^T\beta_{i,j}^{(s)} + (U_g[i,]^{(s)})^TV_g[j,]^{(s)} + \epsilon_{i,j}^*(t)$, where $\epsilon_{i,j}^*(t)$ is a standard normal error; set $\theta_{i,j}^{(s)}(t) = \theta_{i,j}^{(s-1)}(t)$ first, and then replace it by $\theta_{i,j}^*(t)$ with probability $\min\left(\frac{p(Y_{i,j}(t)|\theta_{i,j}^*(t))}{p(Y_{i,j}(t)|\theta_{i,j}^{(s-1)}(t))},1\right)$. 
\end{enumerate}

Here we adopt standard multivariate normal priors for each $\beta_{i,j}$, $u_{i,g}$ and $v_{j,g}$, and the posterior full conditional distributions can be easily derived.

\subsection{Spatial effect estimation}

Although treated as known covariates in parameter estimation, the normalized additive spatial effect functions in Eq.\ref{eq:spatial-func}, $\bar{\xi}_{i}$ and $\bar{\tilde{\xi}}_{i,\text{pos}(j)}$, are not readily available in the optical tracking data and thus need to be estimated. We adopt a different but simpler and more efficient method than Gaussian Markov random field approximation used by \cite{cervone2016multiresolution}, and this method does not require a massive volume of data. 

We first divide the half court $\mathcal{S}$ (47 feet by 50 feet) into $1\,\text{ft} \times 1\,\text{ft}$ tiles and use thin plate splines regression \citep{duchon1977splines} to estimate smooth 2-d functions based on empirical counts on the tiles. For each player $i$, the estimation of $\bar{\xi}_{i}$ is based on the total number of times player $i$ stood on each tile when he made a pass, and for each possible basketball position of his teammate, $\text{pos}(j) \in \{\text{forward (F)},\text{center (C)}, \text{guard (G)}\}$, the estimation of $\bar{\tilde{\xi}}_{i,\text{pos}(j)}$ is based on the total number of times any teammate playing position $\text{pos}(j)$ stood on each tile when he received the ball from the player $i$.

Take the spatial effect function $\bar{\xi}_{i}$ for a certain player $i$ for example. Suppose $\{n_k\}_{k=1}^K$ are the empirical counts of $i$'s passing location on tiles $k=1,\dots,K$ centered at $\{\mathbf{c}_k\}_{k=1}^K$. Set $\tilde{n}_k = \frac{n_k}{\sum_{k=1}^K n_k}$, and the function $\bar{\xi}_{i}: \mathcal{S} \rightarrow \mathbb{R}$ is obtained by minimizing
\begin{equation}
\sum_{k=1}^K\Vert\tilde{n}_k-\bar{\xi}_{i}(\mathbf{c}_k) \Vert^2 + \lambda \int_{\mathcal{S}} \Vert\frac{\partial^2 \bar{\xi}_{i}(\mathbf{s})}{\partial \mathbf{s}^2}\Vert_{F}^2  d\mathbf{s},
\end{equation}
where $\Vert \cdot \Vert_{F}$ is the Frobenius norm of matrices, and the smoothness parameter $\lambda$ is chosen by generalized cross validation, as introduced by \cite{green1993nonparametric}.

\begin{figure}[H]
    \centering
    \begin{subfigure}[b]{0.25\textwidth}
        \vspace*{-0.2in}
        \hspace*{-0.2in}
        \includegraphics[height=1.8in,left]{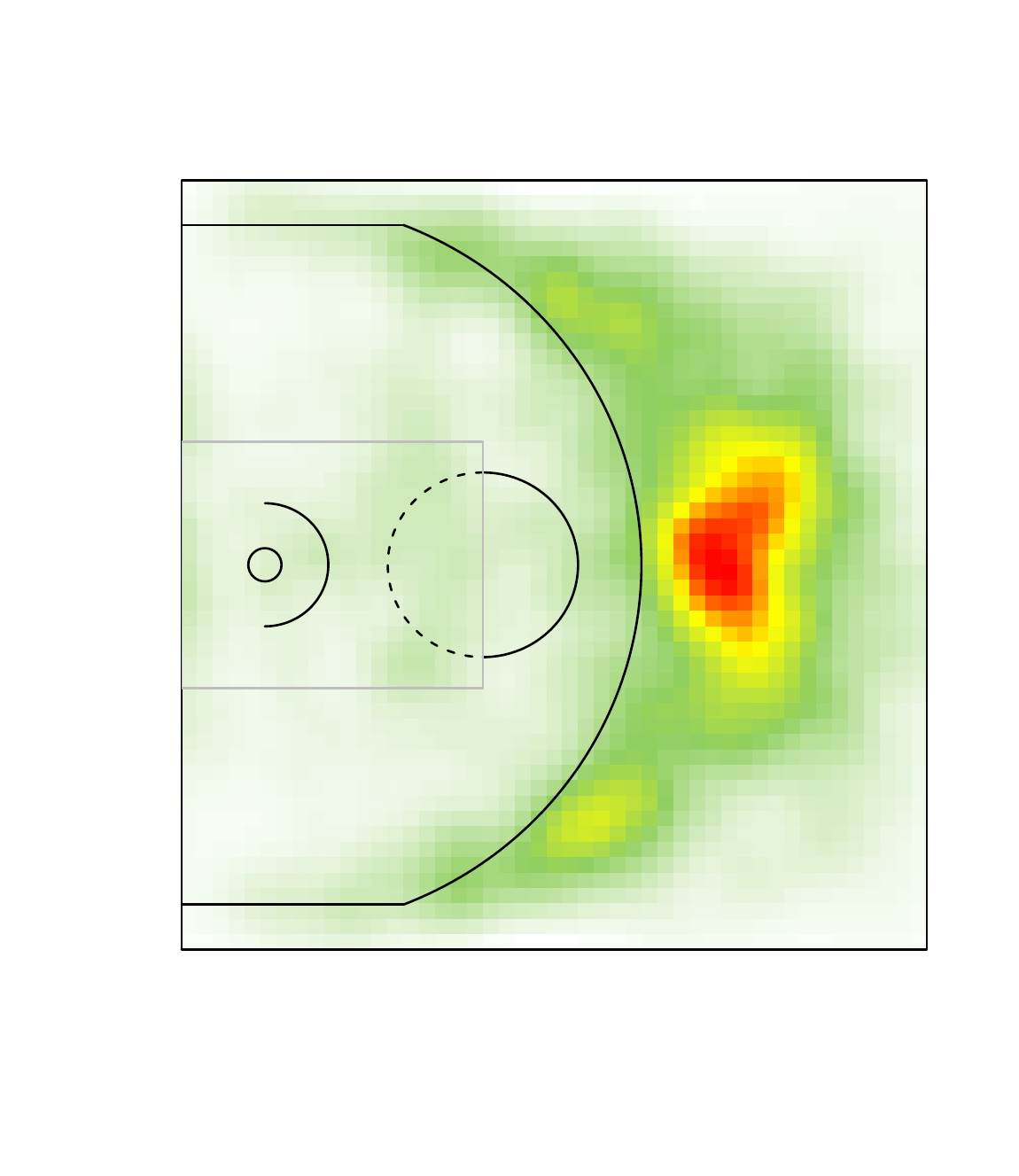}
        \vspace*{-0.3in}
        \caption{$\bar\xi_i$ (make a pass).}
        
    \end{subfigure}%
    ~ 
    \begin{subfigure}[b]{0.25\textwidth}
        \vspace*{-0.2in}
        \hspace*{-0.2in}
        \includegraphics[height=1.8in,left]{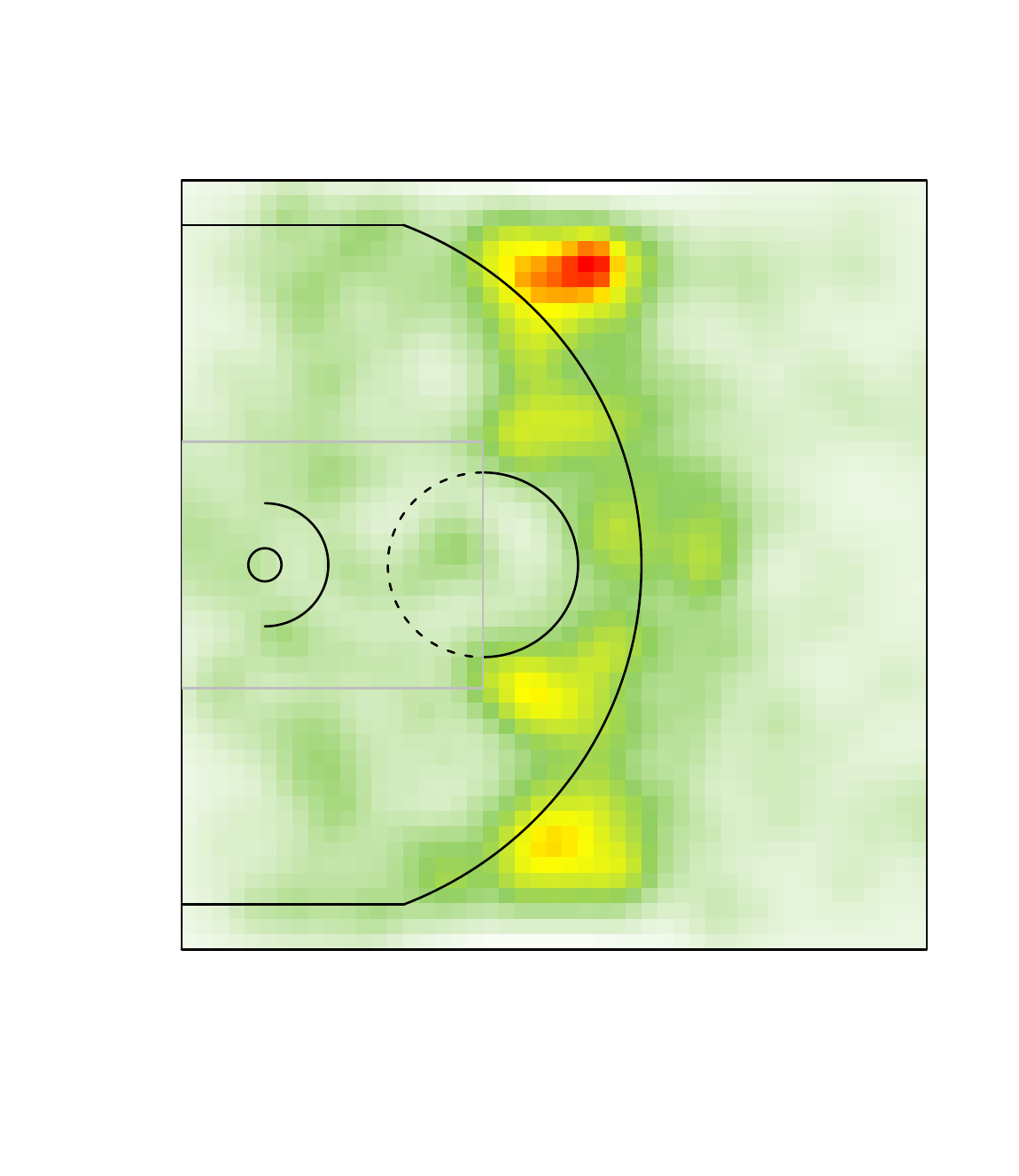}
        \vspace*{-0.3in}
        \caption{$\bar{\tilde{\xi}}_{i,F}$ (pass to F).}
        
    \end{subfigure}%
    ~ 
    \begin{subfigure}[b]{0.25\textwidth}
        \vspace*{-0.2in}
        \hspace*{-0.2in}
        \includegraphics[height=1.8in,left]{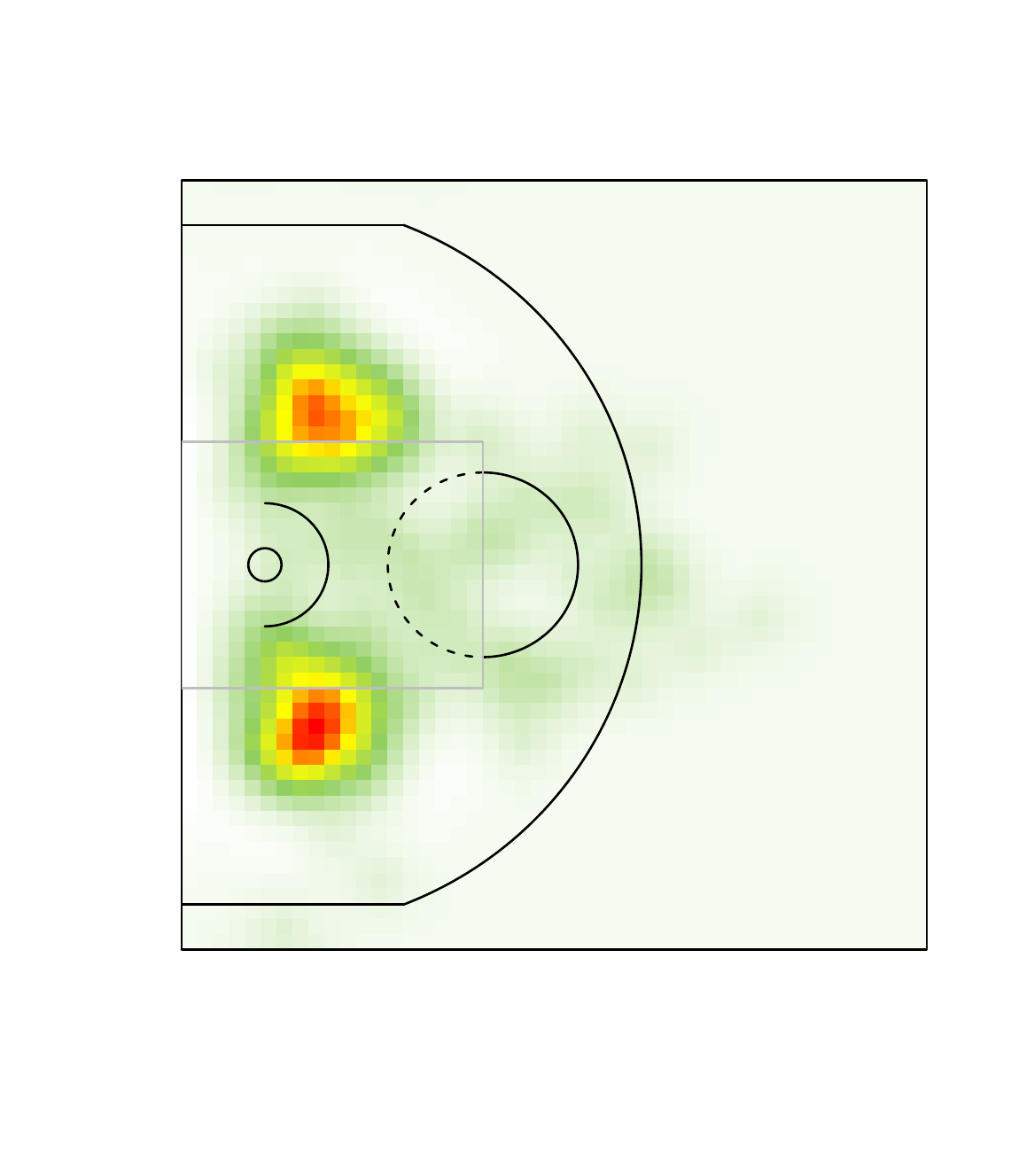}
        \vspace*{-0.3in}
        \caption{$\bar{\tilde{\xi}}_{i,C}$ (pass to C).}
        
    \end{subfigure}%
    ~ 
    \begin{subfigure}[b]{0.25\textwidth}
        \vspace*{-0.2in}
        \hspace*{-0.2in}
        \includegraphics[height=1.8in,left]{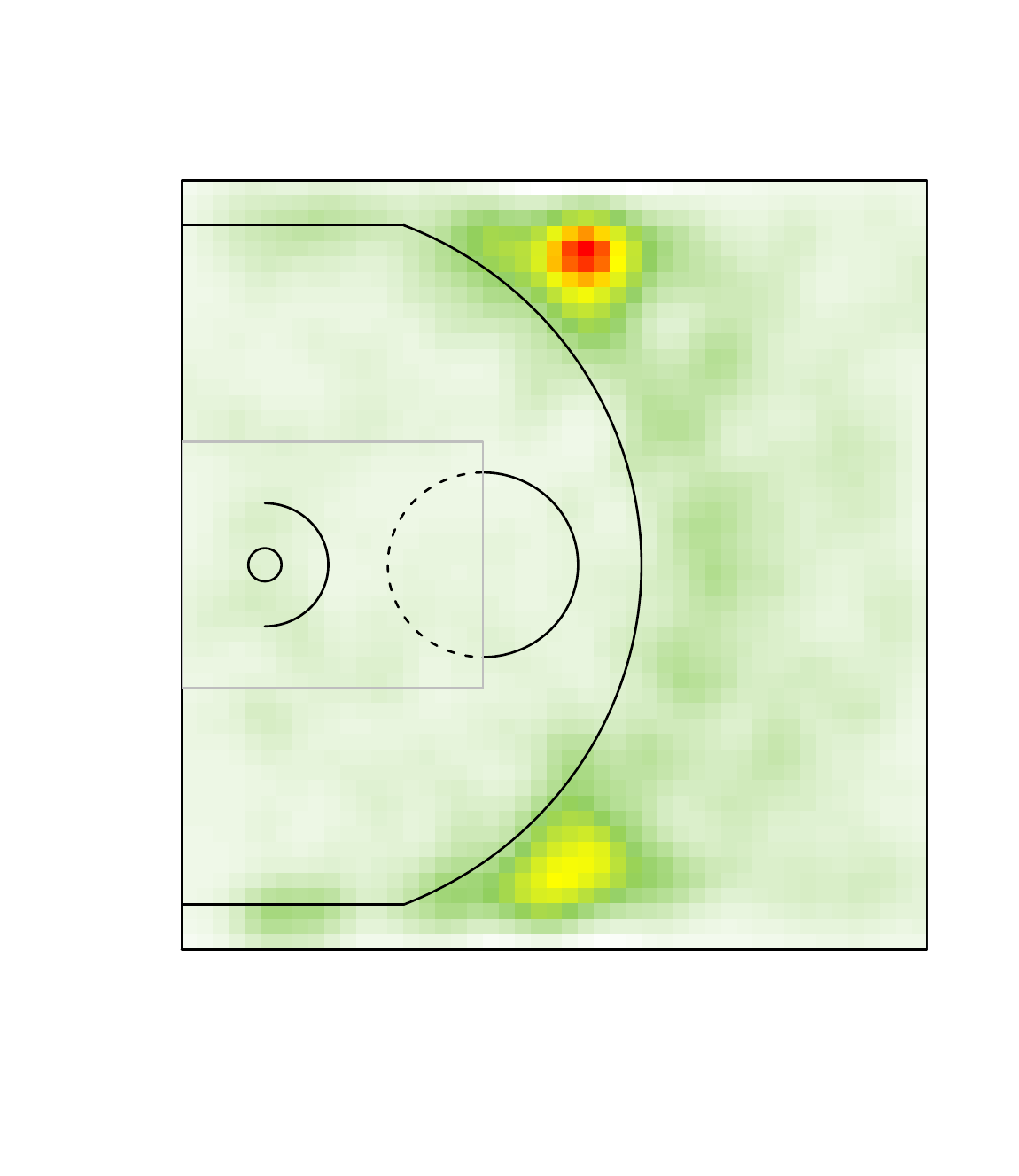}
        \vspace*{-0.3in}
        \caption{$\bar{\tilde{\xi}}_{i,C}$ (pass to G).}
        
    \end{subfigure}
    \caption{Estimated spatial effects $\bar\xi$ and $\bar{\tilde{\xi}}$ for a player $i$ (id number 601140). A redder color corresponds to a higher log-risk of making a pass. For example, this player most frequently passes off the ball when he is just outside the center of the three-point line, and he tends to pass the ball to a center who is approaching the restricted area from either side.}
    \label{fig:spatial-func}
\end{figure}

Figure \ref{fig:spatial-func} visualizes the estimated spatial effect functions for one player 601140, who plays as a guard. As can be observed, there are distinct spatial patterns in his passing choices and preferences, where he tends to pass off the ball right outside the center three-point line (subplot (a)), and his forward teammate(s) is more likely to receive the ball from him when the teammate is at a corner of the three-point line (subplot (b)).

\section{Experiments}

\subsection{Synthetic data}
A synthetic dataset including records in 2 games involving 8 players in total is generated from our model. Around 10,000 observations are generated in total. For each game, the first 90\% of observations are used for model training, and the last 10\% observations are held out as the testing data. Two models, the full model with 2-dimensional multiplicative latent factors(``\textbf{Latent}''), and a subset model with only the time-varying covariates (``\textbf{Covariate}''), are fit on the training dataset. The log-likelihoods of the training data and the held-out data are evaluated for each model.

\begin{figure}[H]
    \centering
    \begin{subfigure}[b]{0.5\textwidth}
        \centering
        \includegraphics[height=1.7in]{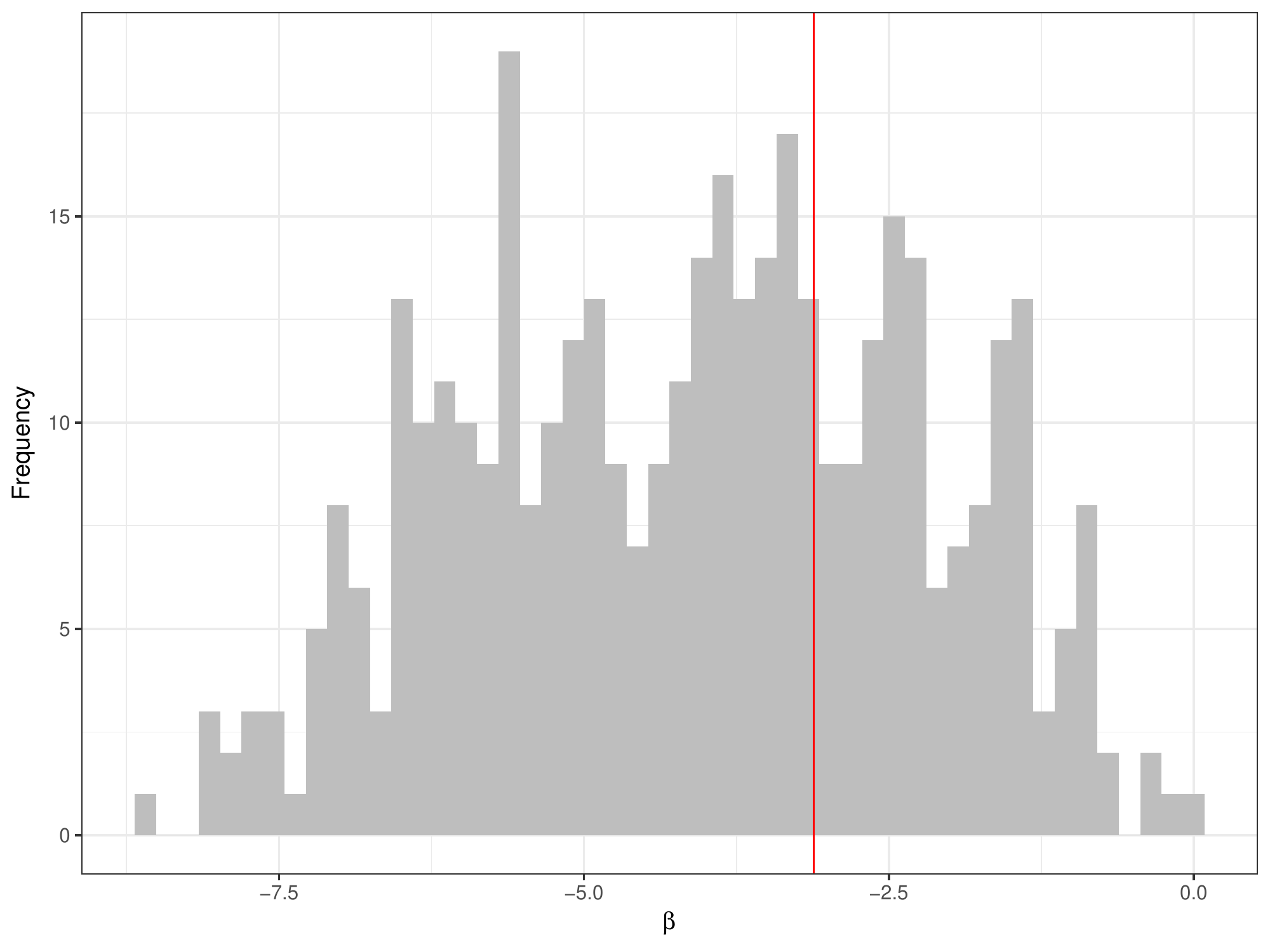}
        \caption{Histogram of posterior samples of $\beta_{i,j,1}$ for a random pair $(i,j)$. The real parameter value is marked by the red vertical line.}
        
    \end{subfigure}%
    ~ 
    \begin{subfigure}[b]{0.5\textwidth}
        \centering
        \includegraphics[height=1.7in]{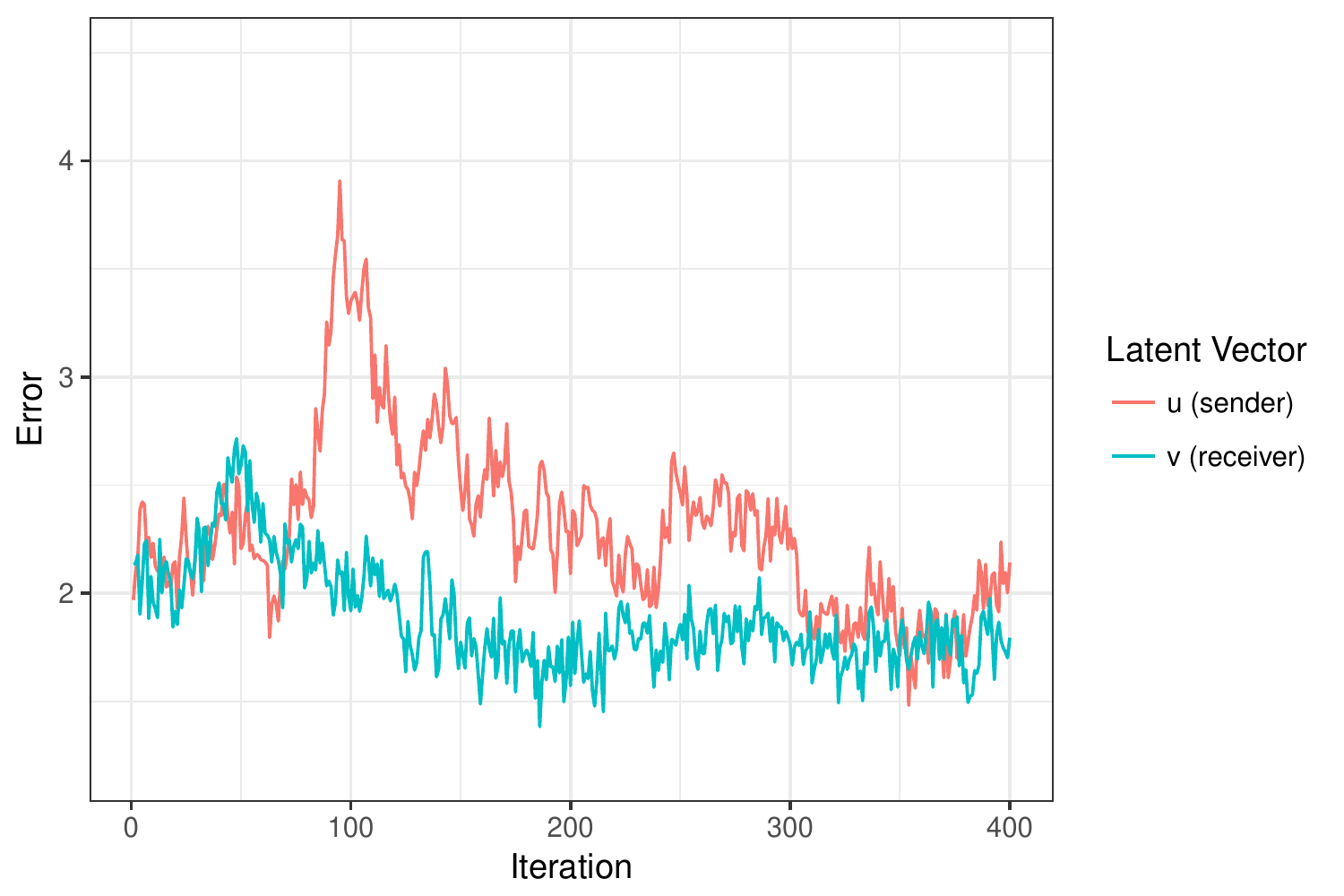}
        \caption{Squared errors of $u_{i,g}$ samples and $v_{i,g}$ samples with respect to the actual latent space vector values for game 1 and a random player $i$.}
       
    \end{subfigure}
    \caption{Parameter recovery checking plots in simulation experiments.}
\label{fig:sim-params}
\end{figure}

Numerical results based on repetitive experiments are presented in Table \ref{tab:sim-lik}. The full model fits the synthetic data better than competitors, and predicts unobserved passing occurrences more accurately. The plots in Figure \ref{fig:sim-params} validate the effectiveness of the sampling scheme (in section 3.2) in recovering the parameters. Plot (a) shows the histogram of the posterior samples of $\beta_{i,j,1}$, the baseline log-risk for player $i$ to pass to $j$, with $i$ and $j$ randomly selected. The posterior samples are approximately centered around the true parameter value (denoted by the red vertical line). Plot (b) visualizes the squared errors (measured by squared euclidean distances) between the samples of sender-specific effect vector and receiver-specific effect vector between the respective true vector values for a random player $i$ in the first game. The squared errors fluctuate in the early iterations due to the random nature of the sampling algorithm, but drop down and stabilize at the end.
\begin{table}[H]
\centering
\caption{Log-likehoods of the full model (``\textbf{Latent}'') and the subset model (``\textbf{Covariate}'') in simulation experiments. The full model constantly outperforms the subset model.}
\vspace{0.1in}
\begin{tabular}{lll}
\toprule
Model & Training log-likelihood & Held-out log-likelihood \\
\midrule
\textbf{Latent} & $-10025.93 \pm 189.31$ & $-1219.80 \pm 57.62$ \\
\textbf{Covariates} & $-10719.68 \pm 120.26$ & $-1314.00 \pm 43.43$ \\
\bottomrule
\end{tabular}
\label{tab:sim-lik}
\end{table}

\subsection{Real data}
\subsubsection{Data description}
The real dataset is collected by SportsVu optical tracking systems from the home arena of a Division I basketball program. This is the first time a public SportsVu dataset for college basketball has been available. Previous NBA SportsVu datasets have been released because every basketball stadium is mandated to retain a tracking video camera. Features were created by taking snapshots of the game every 1/25th of a second and recording each player's location and action, the location of the ball, and general identification information about the game, the teams, and the players \footnote{In the dataset, every player and every team is identified by an id, and every game is identified by the date of the game. To maintain anonymity, we only refer to a player by his randomly assigned id number and leave out the specific date of any game.}. All of the observations were automatically translated and stored into XML files by the SportsVu software, so some of the data were not correctly recorded. The dataset was originally presented as 3 different types of XML files: 

\begin{enumerate}
\item \textbf{Boxscore}:
This dataset shows the overall player statistics (assists, points, rebounds) for each game. It is used as a reference for evaluating player performance.
\item \textbf{Play by Play}: This dataset provides an event summary (such as dribble, foul, pass) for each observation in each game.  It was used to divide the raw data into possessions, which was then converted into individual passing networks.
\item \textbf{Sequence Optical}: This dataset provides the locational summary of each player and of the ball for each game at a $25$Hz resolution. It was used to extract relevant spatio-temporal covariates and map the passing order for each possession.
\end{enumerate}



The final dataset was created by merging the three datasets together by time, player id, and game id. Each game was divided into possessions, which ended on made shots, missed shots, and turnovers. Although the end of a play may not have ended after a non-turnover violation, the locations of the ball and players were reset after these events. For this, non-turnover possessions (i.e. kick ball violation) were also noted as the end of a possession. Possessions that ended with fouls were removed from the dataset to reduce the number of transition states in the model, similar to \citep{cervone2016multiresolution}.     





\subsubsection{Model fitting and results}

We fit the full model (with the dimension of the latent space $R=2$) on the records for all the games from the beginning of December 2014 to the end of January 2015. For model validation, only 90\% of the records in each game are used for model fitting, with the rest 10\% held-out to test model predicting capabilities. Same as in section 4.1, the model is compared with a subset model using only the time-varying covariates. The log-likelihoods on training data and testing data in Table \ref{tab:real-lik} indicate that the addition of multiplicative latent factors yields better explanation of the passing patterns as well as better prediction of passing occurrences in real-time basketball games.

\begin{table}[H]
\vspace*{-0.1in}
\centering
\caption{Log-likehoods of the full model (``\textbf{Latent}'') and the subset model (``\textbf{Covariate}'') on the training dataset and testing dataset.}
\vspace{0.1in}
\begin{tabular}{lll}
\hline
Model & Training log-likelihood & Held-out log-likelihood \\
\hline \hline
\textbf{Latent} & $-679.33 \pm 114.51$ & $-58.52 \pm 11.20$ \\
\textbf{Covariates} & $-917.89 \pm 220.41$ & $-64.68 \pm 12.59$ \\
\hline
\end{tabular}
\label{tab:real-lik}
\end{table}

\begin{figure}[H]
\vspace*{-0.1in}
    \centering
    \begin{subfigure}[b]{0.5\textwidth}
        \centering
        \includegraphics[height=1.7in]{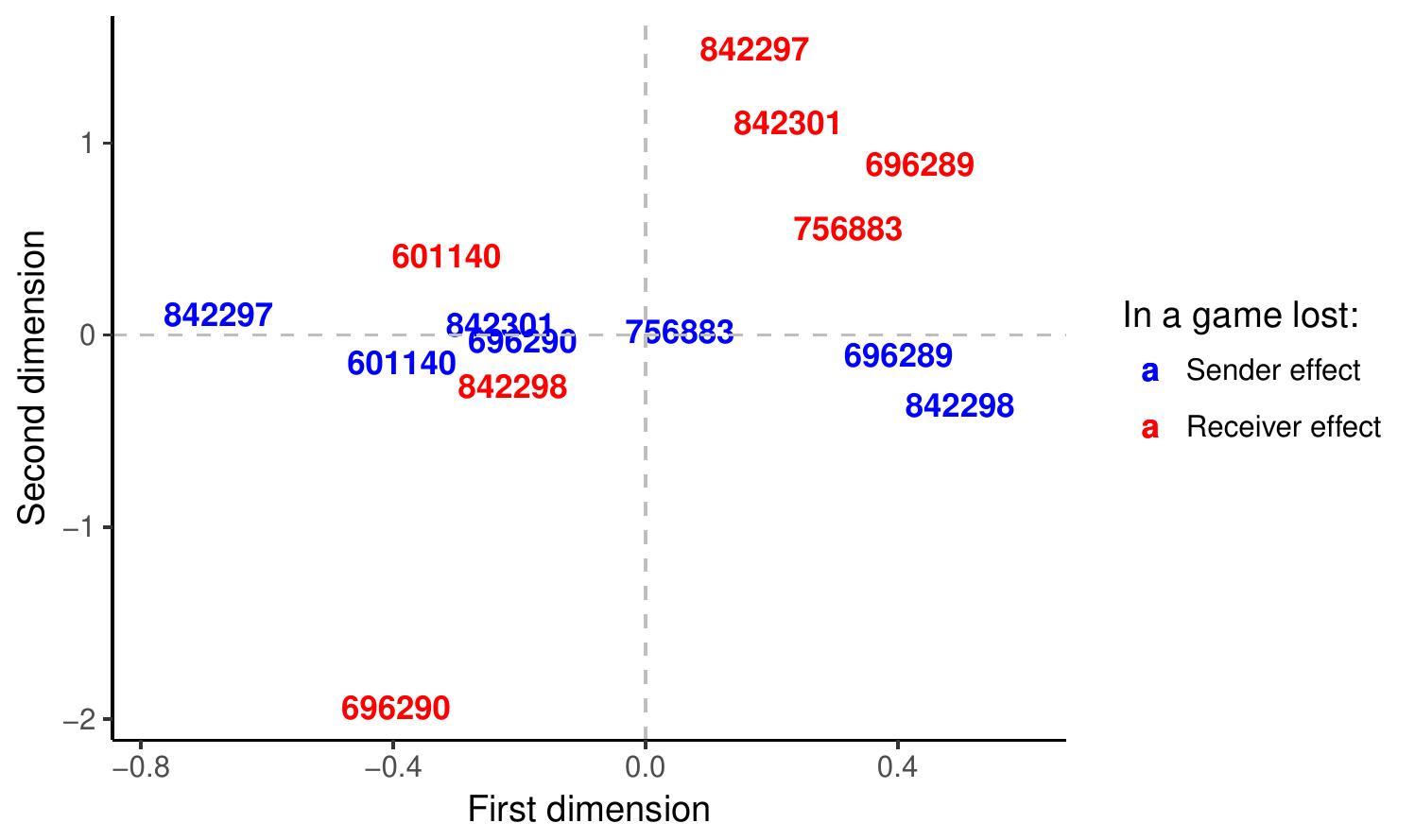}
        \caption{Learned sender-specific effects and receiver specific effects in a lost game.}
        
    \end{subfigure}%
    ~ 
    \begin{subfigure}[b]{0.5\textwidth}
        \centering
        \includegraphics[height=1.7in]{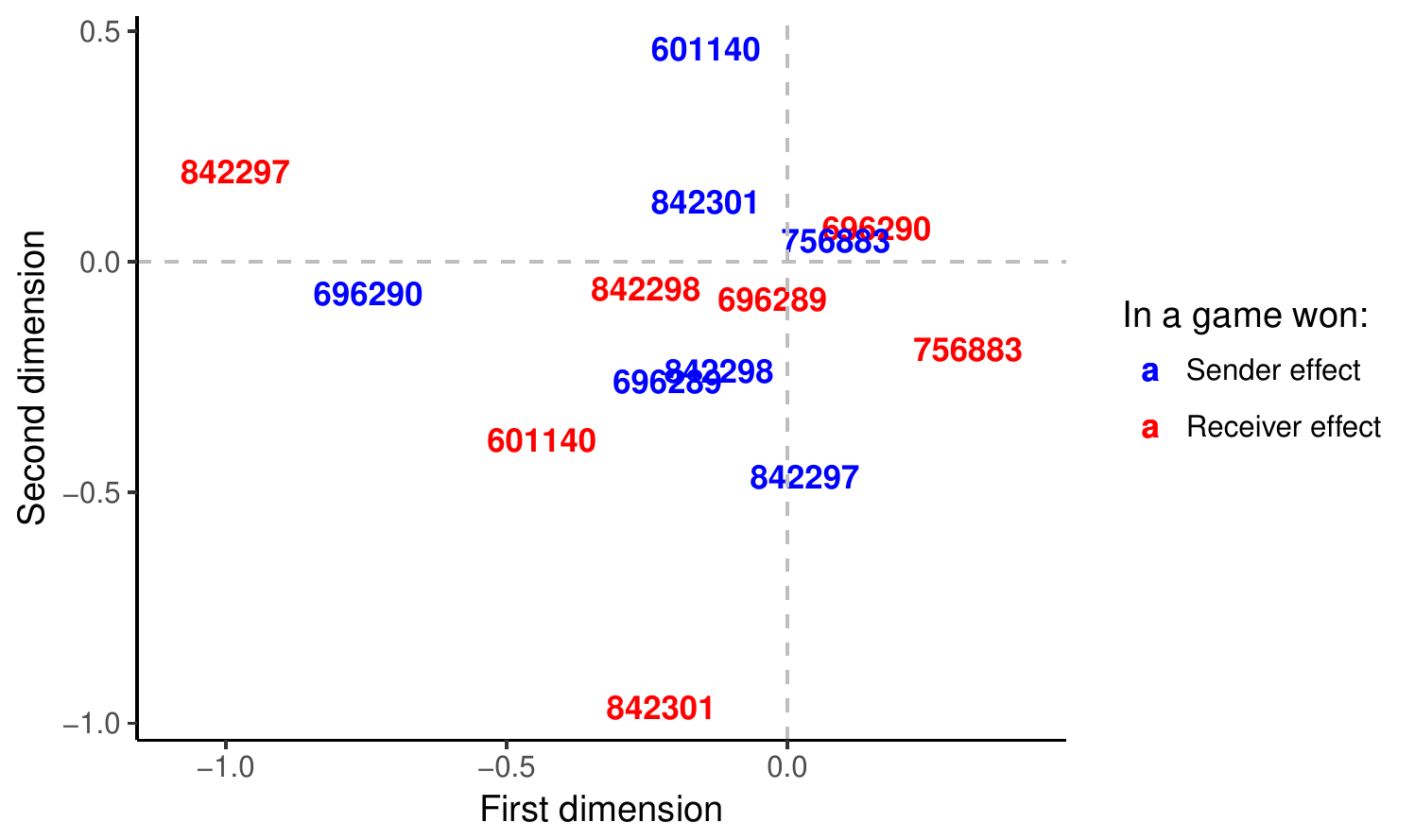}
        \caption{Learned sender-specific effects and receiver specific effects in a successful game.}
        
    \end{subfigure}
    \caption{Passing decision multiplicative latent factors in a loss versus in a victory.}
\label{fig:UV-plot}
\vspace*{-0.1in}
\end{figure}

Figure \ref{fig:UV-plot} offers a visualization of the key results of our model: the sender-specific effects $U_g$ and receiver-specific effects $V_g$ for a game of interest $g$. Each player $i$'s sender-specific effect vector $u_{i,g}$ corresponds to a 2-dimensional coordinate marked by his id number in red, and his receiver-specific effect vector $v_{i,g}$ corresponds to a 2-dimensional coordinate marked by his id number in blue. For player $i$ and $j$, if $u_{i,g}$ and $v_{j,g}$ reside in the same quadrant, then $i$ tends to pass to $j$ more frequently in game $g$. Here we present the sender- and receiver-specific effects for two games, $g_1$ and $g_2$, where the home team lost the former (plot (a)) and won the latter (plot (b)). Players' passing behaviors are apparently different in a loss than in a win. In a victory (game $g_2$), there is more overlap between the sender effects and receiver effects, implying more interactions and more active teamwork. Moreover, in a game lost (game $g_1$), the effects are farther away from the origin than those in a game won (game $g_2$), which indicates that in a loss a player either passes off the ball more frequently or receives the ball more frequently, whereas in a victory the players are more neutral in terms of passing and receiving. 

\section{Conclusion}
We propose social network metrics of basketball game success based on latent factors that capture higher-order patterns in players' passing choices and preferences in basketball games. Our method expands on both the state-of-the-art stochastic process model and latent factor models for binary relational links. Parameter inference is carried out by a Markov chain Monte Carlo sampler, which is effective in recovering the true parameter values, as suggested by experiments on synthetic data. Experiments on the very first high-resolution optical tracking dataset in college basketball show that our model outperforms current state-of-the-art models in both goodness-of-fit and out-of-sample prediction, and that the learned latent sender-specific and receiver-specific effects offer direct interpretation of the interactions among players on the same team and are able to distinguish a win from a loss. While this model is applicable to basketball and other team sports, it can also be translated into modeling longitudinal social networks observations, such as email conversations, international trades, and regional conflicts.

In the future we will include modeling sender-specific and receiver-specific passing effects in each possession rather than each game, considering interactions between defenders and offenders as network links, and scaling up the inference algorithm via sparse matrix/tensor techniques and variational methods.

\newpage

\small
\bibliography{ref}

\end{document}